# Structure and soft magnetic properties of single-domain $Mg_{1-x}Ni_xFe_2O_4$ ($0 \leq x \leq 1.0$) nanocrystals

M. A. Kassem[1,2], A. Abu El-Fadl[1], A. M. Hassan[1], A.M. Gismelssed[3], H. Nakamura[2]

1. Department of Physics, Faculty of Science, Assiut University, Assiut 71516, Egypt,
2. Department of Materials Science and Engineering, Kyoto University, Kyoto 606-8501, Japan,
3. Physics Department, College of Science, Sultan Qaboos University, Muscat, Oman.

**Abstract**

The effects of $Ni^{2+}$ substitution on the structure (lattice parameters, cations distribution, average cations radii, cation-cation/anion bond lengths and angles) and the soft magnetic properties of $Mg_{1-x}Ni_xFe_2O_4$ ($0 \leq x \leq 1.0$) nanocrystals have been studied by x-ray diffraction with Rietveld refinement, transmittance electron microscopy, Mössbauer spectroscopy and magnetization measurements. The mostly inverse spinel structure was found in $MgFe_2O_4$ nanoparticles and the inversion factor is further increased by Ni-substitution up to a complete inversion in $NiFe_2O_4$. The synthesized ferrites with a small particle size (~ 20 – 30 nm) exhibit soft magnetic properties of a single-domain behavior with critical domain size ~ 30 – 40 nm. An observed monotonic increase in magnetization and drastic decrease in coercivity at 5 K by Ni-substitution result in a lowest anisotropy constant $K$u (2.8 kJ/m³) for $x = 0.5$. We discuss the evolution of magnetic properties by Ni-substitution in a correlation with a subsequent cations-redistribution between sites of the two sublattices and corresponding structural changes in the interionic distances.

# 1. Introduction

Synthesis engineering, physics and technological aspects of nanometer-sized magnetic metal oxides have generated a lot of interest due to the huge surface-to-volume ratio that results in tuneable magnetic and electronic properties in these oxides with localized states [1–7]. Among them, spinel magnetic ferrites (SMFs) with a general formula of $MFe_2O_4$, where $M$ stands for a divalent metal ion ($Mg^{2+}$, $Mn^{2+}$, $Co^{2+}$, $Ni^{2+}$, $Cu^{2+}$ and $Zn^{2+}$), are well known by their chemical and thermal stability. Hence, SMFs have attracted much research attention, particularly in nanostructured size, to study their remarkably enhanced electrical, optical, and magnetic properties[8–15]. The spinel structure of bulk SMFs with two sub-lattices of $A$ and $B$ sites can occur in a normal, inverse or mixed spinel structure, where cations in the $A$ and $B$ sites are tetrahedrally and octahedrally coordinated by $O^{2-}$ anions, respectively. In case of normal spinel structure, such as bulk $^{IV}(Zn)^{VI}[Fe]_2O_4$, $M^{2+}$ ions completely locate in the $A$-sites and $Fe^{3+}$ ions at the $B$-sites. While in the inverse spinel structure, such as $^{IV}(Fe)^{VI}[NiFe]_2O_4$, the $A$-sites are completely occupied by a half of $Fe^{3+}$ ions and hence the $B$-site by the other half of $Fe^{3+}$ ions and all of $M^{2+}$ ions.

Another factor that plays an important role in determining the properties of nano-sized SMFs, as well as the surface area does, is the cations distribution between the $A$ and $B$ sites, i.e., the inversion factor ($\delta$) in a spinel structure with general formula ($\{M^{2+}\}_{1-\delta}\{Fe^{3+}\}_\delta)[\{M^{2+}\}_\delta\{Fe^{3+}\}_{2-\delta}]O_4$ [16,17]. Furthermore, substitutions among the $M$ divalent cations ($Mg^{2+}$, $Mn^{2+}$, $Co^{2+}$, $Ni^{2+}$, $Cu^{2+}$ and $Zn^{2+}$) or by a trivalent cation ($Cr^{3+}$, $Mn^{3+}$, $Al^{3+}$, $Ga^{3+}$ etc.) can easily modify the magnetic, optical, (di)electrical and catalytic properties of nanostructured ferrites [18–24]. The origin beyond these enhancements has been explained based on an antistructure model as the cations re-distribution and subsequently emergent active donor or acceptor centers on the large surface of nanosized SMFs[25,26].

Due to their high electrical resistivity, i.e. low eddy current and low dielectric losses[27], nano-sized SMFs with varieties in the spinel structure and tuneable degree of magnetic coupling are candidate materials for vast scope of technological applications including high-density data storage[28], ferrofluids[29], magnetic drug delivery[30], transformers[31], microwave devices[32] and even recently as rare-earth-free permanent magnets[9]. Most of these applications are promoted by the magnetic properties of SMF particles such as the saturation magnetization ($M_s$), coercivity ($H_c$) and hence magnetic anisotropy ($K_u$) which depend on the selected $M^{2+}$ ions and can be tuned by the particles size and shape[9].

$MgFe_2O_4$ is a remarkable ferrimagnetic SMF with magnetic couplings originate only from the magnetic moment of Fe cations distributed between the $A$ and $B$ sites. Due to non-magnetic $Mg^{2+}$ metal ions, the magnetic moment and may be coercivity of $MgFe_2O_4$ are expected to be lower than those of $CoFe_2O_4$ and $NiFe_2O_4$. However, the highest electrical resistivity and lowest dielectric loss among SMFs make soft ferrite $MgFe_2O_4$ is preferred in the high-frequency applications, transformers and low-dissipation magnetic storage[33–35]. On the other hand, NiFe$_2$O$_4$ is a versatile and technologically important soft ferrite material because of its high $M_s$ as well as low $H_c$[36].

In the present work, we study the magnetic properties of the solid-solution between MgFe$_2$O$_4$ and NiFe$_2$O$_4$ in the nanoparticles form by detailed measurements of magnetization and Mössbauer spectroscopy with two objectives in mind. First to accomplish our study of spinel structural properties and cations redistribution with doping in Mg$_{1-x}$Ni$_x$Fe$_2$O$_4$ nanoparticles by further analyses of the XRD results and Mössbauer spectra. Results of XRD, HRTEM, EDX and FTIR, we have previously reported, indicate the successful growth of stoichiometric ferrite nanocrystals of Mg$_{1-x}$Ni$_x$Fe$_2$O$_4$ ($0 \leq x \geq 1.0$) with high crystallinity[37]. The second objective is to examine the correlation trends between the structural parameters, cations distribution and magnetic properties of Mg$_{1-x}$Ni$_x$Fe$_2$O$_4$ nanoparticles. One may manipulate the magnetic properties of a spinel material to meet the demands of some applications.

## 2. Experimental techniques

The Mg$_{1-x}$Ni$_x$Fe$_2$O$_4$ ($0 \leq x \leq 1.0$) nanoparticles used in this study have been synthesised by one-pot microwave combustion as described elsewhere[37]. The synthesized phases were identified by powder x-ray diffraction (XRD) technique using a diffractometer (Philips PW1710, Netherlands) equipped with an automatic divergent slit and CuK$_\alpha$-radiation ($\lambda = 0.15418$ nm) monochromated by a graphite monochromator. The crystal structure parameters have been refined from the observed XRD patterns by a Rietveld analysis method.

The chemical compositions were measured and analyzed by using inductively coupled plasma mass spectrometry (ICP-MS). 10 mg of each powder sample was dissolved in 8 mL of a boiling 37 wt% HCl. After 5 hours on a heater, more diluted HCL up to 10 times in deionized water was added to the digested sample in total volume of 50 mL. Longer time of heating was found to be required to dissolve the Ni-rich samples. Three solution samples with 1000 ppm of Mg, Ni and Fe purchased from Kanto Chemical Co. Inc. were used to prepare reference mixture samples digested in 10-times diluted HCl. A calibration linear curve generated for the reference samples has been used to determine the element concentrations in the digested Mg$_{1-x}$Ni$_x$Fe$_2$O samples in the ICP measurements.

The Mössbauer spectra were recorded at 78 K (liquid nitrogen) with using standard time mode spectrometer (Wiesel) in constant acceleration mode. Mössbauer analysis is used to investigate the magnetic phases and to obtain the Fe distribution in both tetrahedral (*A*)and octahedral (*B*) sites. The radioactive source $^{57}$Co/Rh kept at room temperature was used and the spectrometer was calibrated using *α*-Fe. The Mössbauer spectra were computer-fitted based on Lorentzian lines of equal intensities and line widths. The goodness of fit has been checked by the $\chi^2$-test.

Magnetic properties were investigated by measuring magnetization as function of temperature and field after zero-field-cooling (ZFC) and field-cooling (FC) by using a superconducting quantum interference device (SQUID) of MPMS7 (Quantum Design) in a temperature range of 2 – 300 K and at magnetic fields up to 7 T. The magnetization hysteresis loops, *M* vs. *H* curves, were measured at room temperature and at 5 K after ZTC and FC from room temperature in a magnetic field of 1T.

## 3. Results and discussion
### 3.1. Characterization and XRD analysis

The actual chemical composition of $Mg_{1-x}Ni_xFe_2O_4$ nanoparticles has been investigated by ICP-MS measurements for completely digested samples in diluted HCl as described in section 2. The typical ICP results of atomic ratio with normalization to concentrations of (Mg + Ni + Fe) = 3 are presented in table 1. The observed atomic ratios are very close to nominal compositions.

Figure 1(a) shows the observed XRD patterns of selected $Mg_{1-x}Ni_xFe_2O_4$ samples and results of corresponding Rietveld refinement performed by assuming the cubic spinel structure, space group $Fd\bar{3}m$ (setting #2). No extra peaks other than the shown indexed peaks of spinel cubic structure are observed. The refinement has been performed, by employing the RIETAN-FP system for pattern-fitting structure refinement [38], assuming sites 8a (1/8, 1/8, 1/8) and 16d (½, ½, ½) for atoms lie in the tetrahedral *A*- and octahedral *B*-sites, respectively, while oxygen atoms occupy the 32e site (*x*, *y*, *z*). The ICP actual compositions have been used during the Rietveld analysis and the lattice constant (*a*), oxygen position parameter (*u* = *x* = *y* = *z*), occupancies of ions (*g*) and peaks shape parameters (FWHM, position, intensity, etc.) were refined. Average crystallite size obtained from the Rietveld-estimated broadenings after correction for the internal lattice strain, *ε*, using the well-known Williamson–Hall formula, are about 20 – 30 nm. The estimated crystallite sizes are in agreement with the average particle size observed by TEM indicating successful growth of nano-single crystals[37].

Figure 1(b) shows an example of $MgFe_2O_4$ nanoparticles of almost cubic shape, imaged by TEM, with particle size distribution averaged around 30 nm, shown in one inset. Th high crystallinity is revealed by high-resolution TEM images with lattice fringes structure of adjacent particles, as sown in the top inset. Figure 1(c) shows selected-area electron diffraction (SAED) patterns of $MgFe_2O_4$ nanoparticles show the spotty ring patterns (indicated by 1 to 7), corresponding to the low-angle diffraction lines in XRD, without any additional rings or in-between diffraction spots of secondary phases. The SAED patterns provide a further evidence of the successful synthesis of spinel single-phase nanoparticles with highly crystalline structure. Further details of characterization by TEM, EDX and FTIR are described elsewhere[37].

The values of lattice constant *a* for Mg-Ni ferrites nanoparticles experimentally obtained from XRD by Rietveld refinement are plotted against $Mg^{2+}$ content (*x*) in Fig. 2(a). It is noticeable that the lattice constant linearly decreases with Ni-doping from 8.372 Å to 8.318 Å in a good agreement with Vegard's law [39]. The decrease of lattice constant with increasing $Ni^{2+}$ concentration is attributed to the smaller size of substituent $Ni^{2+}$ ions (0.069 nm) relative to the replaced $Mg^{2+}$ ions (0.072 nm in radius)[40]. The estimated values are consistent with previously observed values for Ni-Mg ferrite synthesized by a co-precipitation method [41].

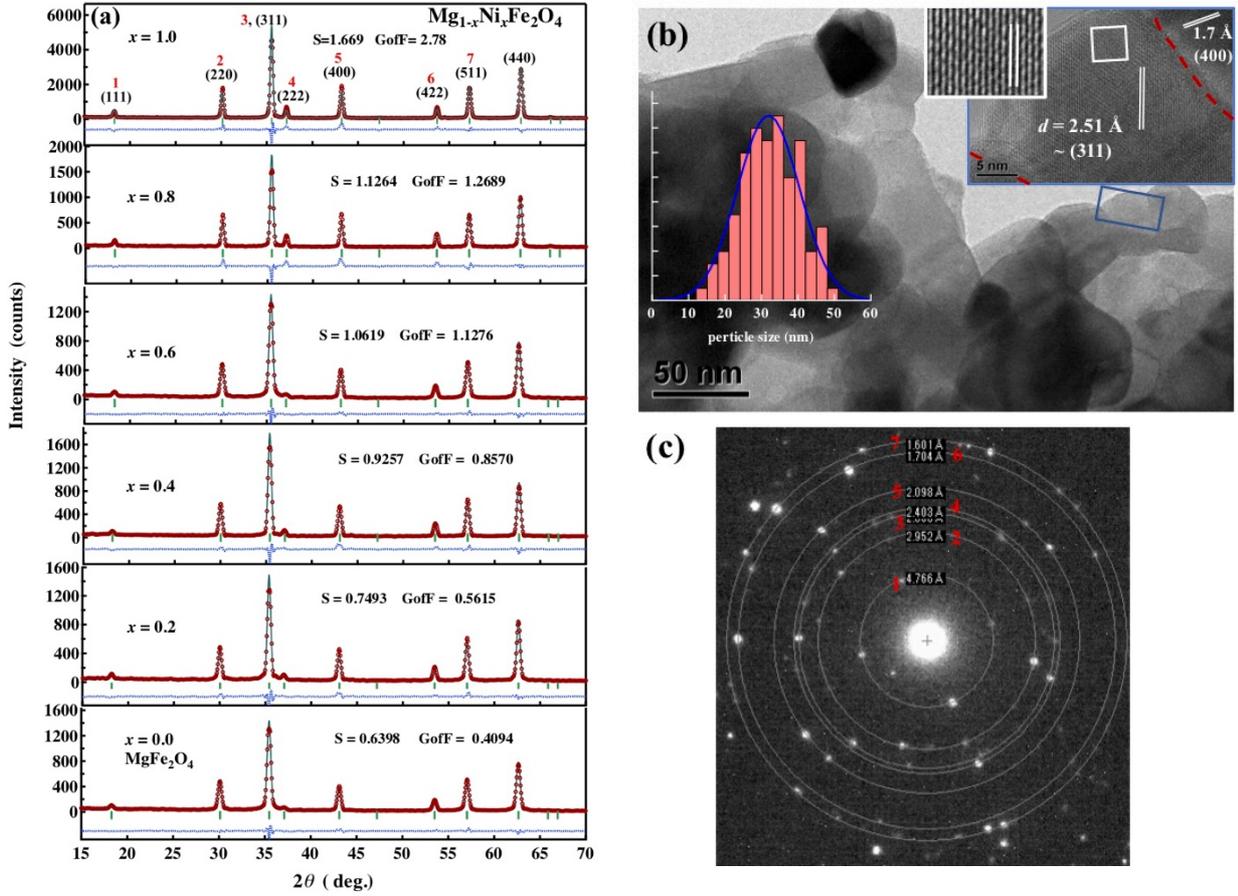

**Figure 1.** **(a)** XRD patterns and Rietveld analysis of $Mg_{1-x}Ni_xFe_2O_4$ nanoparticles at 300 K. **(b)** Example of TEM image for $MgFe_2O_4$ nanoparticles with the particle-size distribution plot, high resolution TEM image of a particle surface-plane and the electron diffraction (ED) patterns are shown as insets. **(c)** and **(d)** show the variation lattice constant $a$ and the inversion factor $\delta$, respectively, with Ni-content $x$.

The free refinement of the ions occupancy $g$ for $MgFe_2O_4$ patterns results in inversion factor $\delta_{XRD} \simeq 0.83$ in a mixed spinel structure with a configuration $(Mg_{0.17}Fe_{0.83})[Mg_{0.415}Fe_{0.585}]_2O_4$. On the other hand, the free refinement of $g$ values in $NiFe_2O_4$ results in $\delta_{XRD} = 0.997$ and reveals the inversed spinel structure in this compound nanoparticles which is in consistence with reported results of nanoparticles with relatively large size[42]. Based on this result and as usually followed with Ni-based ferrites [43], we have assumed the complete occupancy of Ni in the $B$-site while Rietveld refinement of XRD data of the whole series of $Mg_{1-x}Ni_xFe_2O_4$ solid solutions. The change of $\delta_{XRD}$ with Ni content is shown in Fig. 2(b). Interestingly, the refinement results in a mixed spinel structure with $\delta_{XRD} \simeq 0.85$ that is kept almost unchanged with slight Ni-doping before occurrence of a sudden increase to 0.9 at $x \simeq 0.3$ followed by a monotonic increase from 0.9 to 1 with the further substitution by $Ni^{2+}$ ions for $Mg^{2+}$. The cation distribution in $A$ and $B$ sites for all compositions based on XRD results are also listed in table 2. Also, we see from Fig. 2(b) that the trend behavior of $\delta_{XRD}$ with Ni content, estimated from XRD, is in consistence with results of Mössbauer spectroscopy as described below. This means that more $Fe^{3+}$ ions replace $Mg^{2+}$ in the *A*-site resulting in further inversed spinel structure with increasing Ni content above $x \simeq 0.3$ in $Mg_{1-x}Ni_xFe_2O_4$ nanoparticles which should influence the magnetic properties[43–45].

**Table 1:** Composition results of $Mg_{1-x}Ni_xFe_2O_4$ nanoparticles using the inductively coupled plasma mass spectrometry (ICP-MS).

| | ICP-determined atomic ratios with (Mg+Ni+Fe) = 3 | | |
|---|---|---|---|
| $x$ ($Ni^{2+}$) | **Mg** | **Ni** | **Fe** |
| **0** | 0.9927(9) | 0 | 2.007(13) |
| **0.1** | 0.9026(9) | 0.1021(9) | 1.995(14) |
| **0.2** | 0.806(12) | 0.2005(9) | 1.994(19) |
| **0.3** | 0.7070(4) | 0.2911(14) | 2.0018(48) |
| **0.4** | 0.6015(68) | 0.4028(41) | 1.996(10) |
| **0.5** | 0.5098(88) | 0.495(4) | 1.995(17) |
| **0.6** | 0.3925(33) | 0.5943(92) | 2.0131(44) |
| **0.8** | 0.1993(25) | 0.7984(35) | 2.002(13) |
| **0.9** | 0.1042(1) | 0.9059(15) | 1.9898(66) |

The oxygen anions are located on the equipoint 32e ($u$, $u$, $u$) with their positions determined by one parameter $u$. In ideal cubic close-packed spinel structures with the octahedral bond 1.155 times as long as the tetrahedral bond, $u = 0.375$ with assuming the unit cell origin is an $^{IV}A$-site ($\bar{4}3m$ point symmetry) or $u = 0.25$ with assuming the unit cell origin is a $^{VI}B$-site ($\bar{3}m$ point symmetry)[46]. The oxygen position is displaced in the [111] direction and hence the *u*-parameter changes depending on the relative effective radii of the cations in the *A* and *B* sites and lattice volume. In consistence the lattice shrinkage, the refinement results in an oxygen position $u$ that little decreases with the substituent by $Ni^{2+}$ ions from about 0.38 for $MgFe_2O_4$ to 0.376 for $NiFe_2O_4$ nanoparticles, as shown in Fig. 2(b). However, it slightly increases first with low-level doping by $Ni^{2+}$ ions exhibiting a peak at $x \simeq 0.15$ which matches the change in inversion parameter and average cations radii, as described below. The value of u is higher than the value in ideal spinel structure (0.375) for mixed spinels $Mg_{1-x}Ni_xFe_2O_4$ and tends to the ideal value with increasing Ni content.

The preferential occupancy of $Ni^{2+}$ in the octahedral *B*-site results in consequent movement of further $Fe^{3+}$ to occupy the tetrahedral *A*-site which also contains some of the $Mg^{2+}$ ions, i.e. increase in the inversion parameter, by incorporating more Ni ions. This redistribution of the $Fe^{3+}$, $Ni^{2+}$ and $Mg^{2+}$ ions in the *A* and *B* sites is implying a change in the effective ionic radii ($\langle r_A \rangle$ and $\langle r_B \rangle$) of metallic atoms in the two sub-lattices[43]. The variation of $\langle r_A \rangle$ and $\langle r_B \rangle$ in this spinel structure with a general configuration of $(Mg_{1-x-y}Fe_\delta)[Mg_yNi_xFe_{2-\delta}]O_4$, with $0 \leq x \leq 1.0$, $0.8 \geq y \geq 0$ and $\delta = x + y$, can be followed by using their relation to the concentration fraction and radius of each ion in the $A$-site and the $B$-site as,

$$\langle r_A \rangle = (1 - x - y) \cdot r_A(Mg^{2+}) + \delta \cdot r_A(Fe^{3+}), \quad (1)$$

$$\langle r_B \rangle = \frac{y \cdot r_B(Mg^{2+}) + x \cdot r_B(Ni^{2+}) + (2 - \delta) \cdot r_B(Fe^{3+})}{2}, \quad (2)$$

where $r_A(Mg^{2+}) = 0.57$ Å and $r_A(Fe^{3+}) = 0.49$ Å are the ionic radii of $Mg^{2+}$ and $Fe^{3+}$ when are tetragonally coordinated ($A$ site), while $r_B(Mg^{2+}) = 0.72$ Å, $r_B(Ni^{2+}) = 0.69$ Å and $r_B(Fe^{3+}) = 0.645$ Å are ionic radii of $Mg^{2+}$, $Ni^{2+}$ and $Fe^{3+}$ when are octahedrally coordinated ($B$ site), respectively[40]. Figure 2(c) shows the variation of $\langle r_A \rangle$ and $\langle r_B \rangle$ with Ni content. Both $\langle r_A \rangle$ and $\langle r_B \rangle$ decrease with increasing the Ni content. The decrease of $\langle r_A \rangle$ is simply due to the replacement of the larger $Mg^{2+}$ ion (0.57 Å) with a smaller $Fe^{3+}$ (0.49 Å) ions in the tetragonally coordinated A site. The rather nonmonotonic behaviour of $\langle r_B \rangle$ with a kink anomaly at around $x = 0.2$ can be attributed to a competitive occupancy of $Mg^{2+}$ and the smaller $Fe^{3+}$ cations in an octahedral coordination at the *B* site after hosting the incorporated $Ni^{2+}$ cations, with an intermediate size in octahedral coordination, that completely prefer the *B* site. With low-level doing, $Ni^{2+}$ ions go to the *B* site to replace the absent $Mg^{2+}$ ions without change in the inversion degree or the content of $Mg^{2+}$ in the *A* site, as seen in the cations distribution presented in table 2. The size of $Ni^{2+}$ in the octahedral site (0.69 Å) is smaller than that of $Mg^{2+}$ (0.72 Å) resulting in the reduced effective radius $\langle r_B \rangle$ at low Ni content. The rather increase in $\delta$ from 0.8 to 0.9, Fig. 2(b), which means movement of further $Fe^{3+}$ ions from the *B* site with $r_B(Fe^{3+}) = 0.645$ Å to the *A* site with $r_A(Fe^{3+}) = 0.49$ Å, while the vice is versa for some $Mg^{2+}$ ions from $r_A(Mg^{2+}) = 0.57$ Å to $r_B(Mg^{2+}) = 0.72$ Å, results in the observed opposite changes suddenly occur in $\langle r_A \rangle$ and $\langle r_B \rangle$ at $x \simeq 0.2$. The monotonic changes in $\langle r_A \rangle$ and $\langle r_B \rangle$ with further doping are in consistence with behavior of $\delta$ as shown in Fig. 2(a), (b).

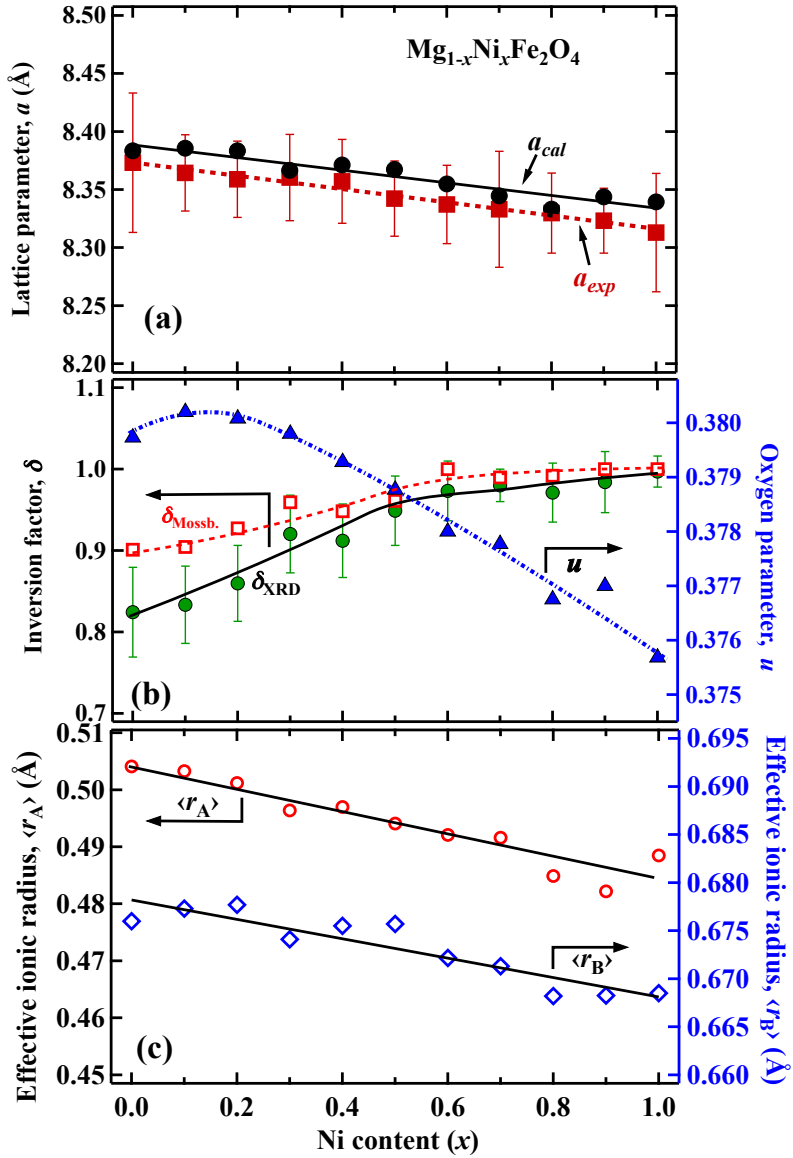

**Figure 2.** Structural variations of $Mg_{1-x}Ni_xFe_2O_4$ nanocrystals with Ni-content $x$: **(a)** observed and calculated lattice constant $a_{exp}$ and $a_{cal}$, solid squares and circles, respectively, **(b)** inversion factor $\delta$ estimated by analyses of XRD patterns (closed circles) and Mössbauer spectra (open squares) and the oxygen position $u$, and **(c)** cations radii $r_A$ and $r_B$. Lines are shown for guidance.

**Table 2:** Structural parameters: crystallite and particle sizes ($d_{xrd}$, $d_{TEM}$), experimental and theoretical lattice parameters ($a_{exp}$, $a_{cal}$), oxygen position parameter ($u$), inversion factor ($\delta$), ionic radii for A-site ($r_A$) and B-site ($r_B$) and chemical formula of $Mg_{1-x}Ni_xFe_2O_4$ nanoparticles indicated by XRD results.

| $x(Ni^{2+})$ | $d_{XRD}/[d_{TEM}]$ (nm) | $a_{exp}$ / Å | $u$ | $\delta_{XRD}$ | $r_A$ (Å) | $r_B$ (Å) | $a_{cal}$ (Å) | Distribution formula $(^{IV}A)[^{VI}B_2]O_4$ |
|---|---|---|---|---|---|---|---|---|
| 0   | 24/[29]    | 8.360(70) | 0.3797 | 0.827(70) | 0.5038 | 0.6760 | 8.3831 | $(Mg_{0.176}Fe_{0.824})[Mg_{0.813}Fe_{1.176}]O_4$ |
| 0.1 | 20.72      | 8.354(33) | 0.3802 | 0.829(30) | 0.5037 | 0.6746 | 8.3791 | $(Mg_{0.167}Fe_{0.833})[Ni_{0.102}Mg_{0.738}Fe_{1.167}]O_4$ |
| 0.2 | 20.69/[27] | 8.352(32) | 0.3801 | 0.848(38) | 0.5022 | 0.6738 | 8.3746 | $(Mg_{0.14}Fe_{0.86})[Ni_{0.201}Mg_{0.668}Fe_{1.114}]O_4$ |
| 0.3 | 25.41      | 8.346(35) | 0.3798 | 0.902(45) | 0.4979 | 0.6743 | 8.3694 | $(Mg_{0.08}Fe_{0.92})[Ni_{0.291}Mg_{0.627}Fe_{1.08}]O_4$ |
| 0.4 | 21.27/[27] | 8.346(37) | 0.3793 | 0.922(35) | 0.4974 | 0.6729 | 8.3648 | $(Mg_{0.088}Fe_{0.912})[Ni_{0.404}Mg_{0.515}Fe_{1.088}]O_4$ |
| 0.5 | 20.29      | 8.335(34) | 0.3788 | 0.936(41) | 0.4948 | 0.6728 | 8.3605 | $(Mg_{0.051}Fe_{0.949})[Ni_{0.496}Mg_{0.46}Fe_{1.051}]O_4$ |
| 0.6 | 22.73/[29] | 8.338(35) | 0.3780 | 0.955(42) | 0.4930 | 0.6722 | 8.3562 | $(Mg_{0.027}Fe_{0.973})[Ni_{0.59}Mg_{0.363}Fe_{1.027}]O_4$ |
| 0.7 | 22.73      | 8.333(55) | 0.3778 | 0.992(18) | 0.4916 | 0.6713 | 8.3516 | $(Mg_{0.012}Fe_{0.988})[Ni_{0.7}Mg_{0.28}Fe_{1.012}]O_4$ |
| 0.8 | 25.68/[22] | 8.330(34) | 0.3768 | 0.971(42) | 0.4924 | 0.6694 | 8.3478 | $(Mg_{0.029}Fe_{0.971})[Ni_{0.798}Mg_{0.17}Fe_{1.029}]O_4$ |
| 0.9 | 24.85      | 8.323(28) | 0.3770 | 0.982(37) | 0.4916 | 0.6683 | 8.3436 | $(Mg_{0.016}Fe_{0.984})[Ni_{0.911}Mg_{0.089}Fe_{1.016}]O_4$ |
| 1   | 29/[31]    | 8.313(51) | 0.3757 | 0.997(19) | 0.4902 | 0.6674 | 8.3392 | $(Fe)[NiFe]O_4$ |

To examine the estimation reliability of the effective ionic radii, one can estimate a theoretical lattice constant ($a_{cal}$) and compare to the experimentally observed values from XRD refinements. The lattice parameter can be calculated as[47],

$$a_{cal} = \frac{8(\langle r_A \rangle + r_O) + 8\sqrt{3}(\langle r_B \rangle + r_O)}{3\sqrt{3}}, \tag{3}$$

where $r_O$ represents the radius of the oxygen ion (1.38 Å)[40]. The calculated values are presented in table 2 and together with the experimental values in Fig. 2(a). The calculated values are generally higher than the experimental ones which may indicate slightly overestimated values of $\langle r_A \rangle$ and $\langle r_B \rangle$ or may be due to a used value of $r_O$ slightly high than actual value. However, it is observed in Fig. 2(a) that the theoretically calculated lattice parameter $a_{cal}$ follows the same trend as that experimentally obtained $a_{exp}$ indicating correctly estimated evolution with Ni doping. All structure parameters derived from the Rietveld analysis are presented in table 2 and Fig. 2 as functions for the Ni-content.

In Néel's molecular field theory, the magnetic properties of the completely localized ferrites mainly depend on the strength of magnetic interaction between atoms in the crystallography different $A$ and $B$ sites as well as do on the cations distribution[48]. The anti-parallel A-B magnetic interaction is the strongest in a ferrimagnetic material and both B-B and A-A interaction are following, although all tend to be negative. Further, an indirect magnetic interaction between metal ions through the oxygen anions, i.e. superexchange mechanism, is expected in a ferrimagnetic similar to that in an antiferromagnetic material[48]. As the strengths of these magnetic exchanges and superexchanges are decided by the inter-ionic distances and angles (proportional inversely to lengths and directly to angles), we have investigated the cation-cation (Me-Me), cation-anion (Me-O) bond lengths and the Me-O-Me angles. Figure 3 shows the unique Me-Me ($b$, $c$, $d$, $e$ and $f$) and Me-O ($p$, $q$, $r$, $s$) bonds and Me-O-Me ($\theta_1$, $\theta_2$, $\theta_3$, $\theta_4$, $\theta_5$) angles in the spinel $AB_2O_4$ structure established with VESTA software[49]. The bond lengths and angles of $Mg_{1-x}Ni_xFe_2O_4$ nanoparticle system can be calculated from the XRD refinement results of lattice parameter $a$ and oxygen parameter $u$ by using equations proposed by Lakhani et al.[47]. The values well agree with values obtained by employing VESTA software after modifying the structure with structural parameters and cations distribution obtained from XRD results by Rietveld refinement. The bond lengths and angles are listed in tables 3 and 4. It is observed from table 3 that all the bond lengths of Me-Me and Me-O, except a slight increase in $p$, with Ni doping which imply an increase in the strength of all possible direct and indirect (through oxygen) magnetic interactions. On the other hand, the A-O-B angles ($\theta_1$ and $\theta_2$) and A-O-A angle ($\theta_5$) increase implying a strengthen exchange (negative) between the magnetic atoms in the two $A$ and $B$ sublattice and also in the $A$ sublattice (positive) with increasing Ni content. An observed slight decrease in the B-O-B angles ($\theta_3$, $\theta_4$) may indicate weakening of the B-B interaction, however, enhancement in the exchange integral $J_{AB}/J_{BB}$ is expected by Ni substitution for Mg in the $Mg_{1-x}Ni_xFe_2O_4$ system.

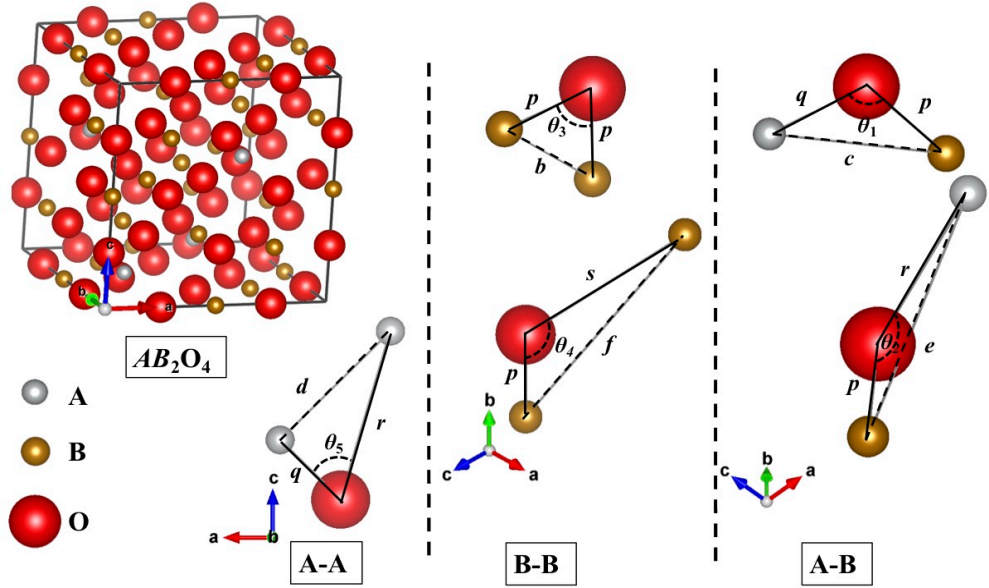

**Figure 3.** The unique Me-Me and Me-O bonds and Me-O-Me angles in the spinel ferrite crystal structure with the tetrahedral $A$ site at 8a(1/8, 1/8, 1/8) and octahedral $B$ site at 16d(½, ½, ½).

**Table 3**: Interionic Me-Me ($b$, $c$, $d$, $e$ and $f$) and Me-O ($p$, $q$, $r$, $s$) distances and the octahedral to tetrahedral bond ration $\gamma = p/q$, estimated from the XRD and Rietveld results

| | cation-cation (Å) | | | | | cation-anion (Å) | | | | |
|---|---|---|---|---|---|---|---|---|---|---|
| $x(Ni^{2+})$ | $b$ (B-B) | $c$ (A-B) | $d$ (A-A) | $e$ (A-B) | $f$ (B-B) | $p$ (B-O) | $q$ (A-O) | $r$ (A-O) | $s$ (B-O) | $\gamma$ |
| 0 | 2.9558 | 3.4660 | 3.6201 | 5.4301 | 5.1196 | 2.0505 | 1.8785 | 3.5971 | 3.6429 | 1.09 |
| 0.1 | 2.9535 | 3.4633 | 3.6173 | 5.4260 | 5.1157 | 2.0450 | 1.8839 | 3.6074 | 3.6424 | 1.09 |
| 0.2 | 2.9530 | 3.4628 | 3.6167 | 5.4251 | 5.1148 | 2.0457 | 1.8819 | 3.6035 | 3.6412 | 1.09 |
| 0.3 | 2.9507 | 3.4601 | 3.6139 | 5.4209 | 5.1108 | 2.0464 | 1.8763 | 3.5929 | 3.6370 | 1.09 |
| 0.4 | 2.9508 | 3.4601 | 3.6140 | 5.4210 | 5.1109 | 2.0508 | 1.8689 | 3.5786 | 3.6346 | 1.10 |
| 0.5 | 2.9467 | 3.4553 | 3.6090 | 5.4134 | 5.1038 | 2.0522 | 1.8589 | 3.5595 | 3.6271 | 1.10 |
| 0.6 | 2.9479 | 3.4567 | 3.6105 | 5.4157 | 5.1060 | 2.0595 | 1.8486 | 3.5397 | 3.6249 | 1.11 |
| 0.7 | 2.9461 | 3.4546 | 3.6082 | 5.4123 | 5.1028 | 2.0601 | 1.8441 | 3.5312 | 3.6216 | 1.12 |
| 0.8 | 2.9451 | 3.4534 | 3.6070 | 5.4105 | 5.1011 | 2.0679 | 1.8287 | 3.5018 | 3.6154 | 1.13 |
| 0.9 | 2.9426 | 3.4505 | 3.6040 | 5.4059 | 5.0968 | 2.0806 | 1.8023 | 3.4512 | 3.6041 | 1.15 |
| 1 | 2.9391 | 3.4464 | 3.5996 | 5.3995 | 5.0907 | 2.0726 | 1.8096 | 3.4651 | 3.6029 | 1.15 |

As the oxygen parameter $u$ does, the octahedral to tetrahedral bond ration $\gamma = p/q$ increases and tends the ideal close-packed spinel structure value (1.155), it increases from about 1.1 in $MgFe_2O_4$ to 1.15 in $NiFe_2O_4$.

**Table 4**: Estimated values of the cation-anion-cation angles with Ni-content

| | cation-anion-cation (deg.) | | | | |
|---|---|---|---|---|---|
| $x$ (Ni$^{2+}$) | $\theta_1$ (A-O-B) | $\theta_2$ (A-O-B) | $\theta_3$ (B-O-B) | $\theta_4$ (B-O-B) | $\theta_5$ (A-O-A) |
| 0 | 123.744 | 146.790 | 92.231 | 125.779 | 75.590 |
| 0.1 | 123.595 | 146.097 | 92.460 | 125.830 | 75.177 |
| 0.2 | 123.633 | 146.273 | 92.401 | 125.817 | 75.282 |
| 0.3 | 123.722 | 146.686 | 92.265 | 125.787 | 75.528 |
| 0.4 | 123.888 | 147.467 | 92.013 | 125.730 | 75.990 |
| 0.5 | 124.051 | 148.250 | 91.767 | 125.674 | 76.447 |
| 0.6 | 124.297 | 149.470 | 91.399 | 125.590 | 77.146 |
| 0.7 | 124.371 | 149.844 | 91.290 | 125.565 | 77.357 |
| 0.8 | 124.699 | 151.557 | 90.810 | 125.454 | 78.304 |
| 0.9 | 124.619 | 151.128 | 90.927 | 125.481 | 78.070 |
| 1 | 125.044 | 153.471 | 90.312 | 125.338 | 79.318 |

### 3.2. Mössbauer spectroscopy and cations distribution

Figures 4(a) displays the Mössbauer spectra of selected compositions of Mg$_{1-x}$Ni$_x$Fe$_2$O$_4$ nanoparticles measured at 78 K. The dots in the figures represent the experimental data and the fine lines through the data points are the fitting lines of the spectra. The estimated hyperfine parameters extracted from the spectral fitting; the isomer shift (*IS*), quadrupole splitting ($\Delta$), magnetic hyperfine field ($H_{hf}$), peak width ($\Gamma$) and the indicated cations distribution between ($^{IV}A$) and [$^{VI}B$], all are presented in table 5.

The analysis of the spectra exhibit two well-resolved Zeeman sextets assigned to the Fe$^{3+}$ ions in the tetrahedral ($^{IV}A$) and octahedral ($^{VI}B$) sublattices which are typical of ferrimagnetic behaviour. The sextet assigned to the Fe$^{3+}$ ions at the $^{VI}B$-site exhibits a larger hyperfine field compared to the sextet of Fe$^{3+}$ ions occupying the $^{IV}A$-site as previously observed for SMFs [50–52]. The values of both *IS* and $H_{hf}$ are consistent with the high spin $d^5$ (Fe$^{3+}$) state with ranges of *IS* ≈ 0.05 – 0.6 mm s$^{-1}$ and $H_{hf}$ ≈ 22 ⟨*Sz*⟩, where ⟨*Sz*⟩= 5/2[50,53]. Figures 4(b) and (c) show the Ni-content dependence of the estimated Isomer shift (*IS*) and the hyperfine field ($H_f$). It is clear that *IS* for the $^{VI}B$-site have values larger than those for the $^{IV}A$-site which can be attributed to longer B-O$^{2-}$ bonds compared to A-O$^{2-}$ bonds, $\gamma = p/q > 1$, as shown in table 3. The change in bond lengths in table 3 elucidate the slight decrease of *IS* for the $^{IV}A$-site and slight increase for the $^{VI}B$-site with Ni substitution, which is observed in the general behaviour of *IS* of each site. The $^{IV}A$-site sextet line width is observed less than that of the $^{VI}B$-site sextet due to the presence of multiple hyperfine fields at the octahedral Fe$^{3+}$ nuclei caused by the three Me ions (Fe$^{3+}$, Ni$^{2+}$ and Mg$^{2+}$) in the B-site [54]. It is also clear from Fig. 4(c) and table 5 that $H_{hf}$ increases at both *A* and *B* sites, in general behaviors, while the nonmagnetic Mg$^{2+}$ are replaced by the magnetic Ni$^{2+}$, which can be explained by possibly added positive transfer field from the further incorporated $Ni^{2+}$ ions. Furthermore, the quadrupole splitting is insignificant with values close to zero in both *A* and *B* sites (table 5) as previously observed in the presence of strong magnetic interaction, where the quadrupole interactions distribution arising from chemical disorder produces an appreciable broadening of the individual Zeeman lines without observable quadrupole line shifts [51].

The concentrations of $Fe^{3+}$ ions at the *B* and *A*-sites are estimated from the relative areas of $^{IV}A$ and $^{VI}B$ sextets and the corresponding inversion factor $\delta_{Moss}$ values are listed in table 5. By assuming that $Ni^{2+}$ ions prefer to occupy the octahedral *B*-sites, a cation distribution is estimated for each composition and also described in table 4. As seen in Fig. 2(b), the cations distribution suggested by Mössbauer spectroscopy is consistent with the $\delta_{XRD}$ results obtained from XRD refinement, particularly at high Ni concentrations and within the uncertainty levels shown in table 2.

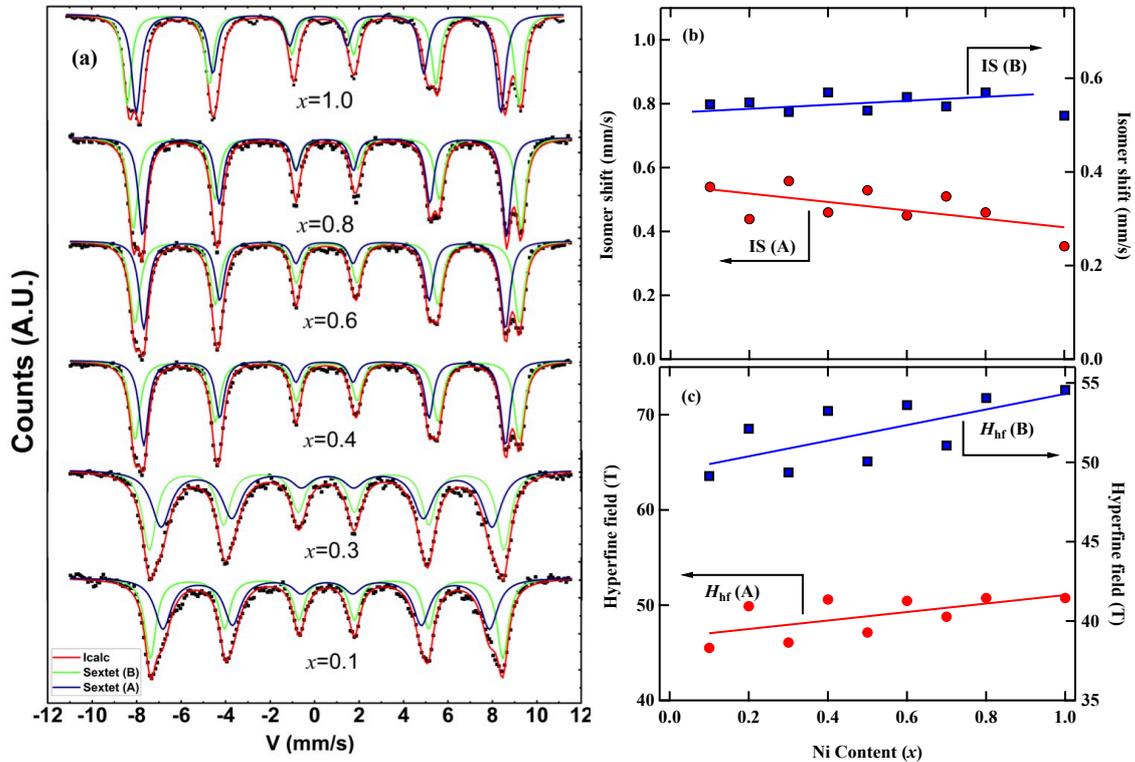

**Figure 4:** (a) Mössbauer spectra of $Mg_{1-x}Ni_xFe_2O_4$ nanoparticles at 78 K. (b) and (c) show the Ni-content dependence of the estimated Isomer shift and hyperfine magnetic field, respectively.

### 3.3. Magnetic Properties

Figure 5 shows the temperature dependence of the magnetization *M*(*T*) measured at different magnetic fields after field-cooling (FC) and zero-field-cooling (ZFC). Both FC and ZFC magnetization *M*(*T*) curves of the two end-members, $MgFe_2O_4$ and $NiFe_2O_4$, are shown in Figs. 5(a) and (b), respectively. It is clearly observed that the *M*(*T*) measured at low fields of 0.01 and 0.02 T after a ZFC process bifurcates from the FC curve, while *M*(*T*) is almost reversable after FC and ZFC at high magnetic field of 1 T. A peak in the ZFC magnetization is expected with the exact position of the peak depending on the applied field which implies a possible spins blocking in ferrimagnetic particles with size comparable to a single domain size[37,55–57]. To examine the blocking behavior in the ZFC mode, we have measured the magnetization after ZFC at different fields below 1T. The ZFC data are shown in Figs. 5(c) and (d) for MgFe$_2$O$_4$ and NiFe$_2$O$_4$, respectively. The peak in the ZFC magnetization is evident in each curve with the exact peak (blocking) temperature ($T_p$) that decreases with increasing the applied field. The insets of Figs. 5(c) and (d) show the field dependence of the $T_p$ for MgFe$_2$O$_4$ and NiFe$_2$O$_4$, respectively, with $T_p$ drastically decreases with increasing field at the low-field region ($H \lesssim 1$ kOe) and slightly decrease at higher fields. As mostly observed for SMF nanoparticles, the ZFC $M(T)$ behaviors of Mg-Ni ferrites with broad peaks look due to strong interparticle interactions where particles are not enough spatially separated rather than superparamagnetic core-shell state[56]. The $M(T)$ of NiFe$_2$O$_4$ indicate stronger interparticle interactions relative to MgFe$_2$O$_4$.

**Table 5:** Mossbauer hyperfine parameters; the isomer shift (*IS*), quadrupole splitting (*Δ*), magnetic hyperfine field ($H_{hf}$) and peak width (*Γ*) as well as corresponding inversion factor ($\delta_{Moss}$) of Mg$_{1-x}$Ni$_x$Fe$_2$O$_4$ nanoparticles

| $x$(Ni$^{2+}$) | Site | *IS* (mm/s) | *Δ* (mm/s) | $H_{hf}$ (T) | *Γ* (mm/s) | Area (%) | $\delta_{Moss}$ |
|---|---|---|---|---|---|---|---|
| **0.1** | Sextet (A) | 0.54 | -0.017 | 45.52 | 0.45 | 45.22 | 0.905 |
|  | Sextet [B] | 0.5444 | 0.0001 | 49.139 | 0.304 | 54.78 |  |
| **0.2** | Sextet (A) | 0.4389 | -0.0065 | 49.894 | 0.3 | 46.37 | 0.927 |
|  | Sextet [B] | 0.5485 | 0.015 | 52.12 | 0.33 | 53.63 |  |
| **0.3** | Sextet (A) | 0.558 | -0.023 | 46.08 | 0.52 | 47.97 | 0.959 |
|  | Sextet [B] | 0.528 | 0.008 | 49.37 | 0.368 | 52.03 |  |
| **0.4** | Sextet (A) | 0.46 | -0.0068 | 50.605 | 0.3 | 47.41 | 0.948 |
|  | Sextet [B] | 0.57 | 0.0129 | 53.244 | 0.309 | 52.59 |  |
| **0.5** | Sextet (A) | 0.531 | -0.005 | 50.06 | 0.29 | 48.05 | 0.961 |
|  | Sextet [B] | 0.529 | -0.013 | 47.14 | 0.41 | 51.95 |  |
| **0.6** | Sextet (A) | 0.45 | 0.0044 | 50.45 | 0.25 | 47.73 | 1 |
|  | Sextet [B] | 0.56 | 0.0112 | 53.613 | 0.27 | 52.27 |  |
| **0.7** | Sextet (A) | 0.51 | -0.025 | 48.79 | 0.378 | 48.8 | 0.990 |
|  | Sextet [B] | 0.54 | 0.0054 | 51.065 | 0.279 | 51.2 |  |
| **0.8** | Sextet (A) | 0.46 | 0.0006 | 50.757 | 0.23 | 49.6 | 0.992 |
|  | Sextet [B] | 0.57 | 0 | 54.058 | 0.251 | 50.4 |  |
| **1.0** | Sextet (A) | 0.354 | 0.005 | 50.75 | 0.50 | 50 | 1 |
|  | Sextet [B] | 0.52 | 0.008 | 54.56 | 0.47 | 50 |  |

To study the magnetic properties of $Mg_{1-x}Ni_xFe_2O_4$ nanoparticles as soft SMFs, the magnetization curves $M(H)$ have been measured for the whole series at a low temperature, 5 K, as well as at room temperature, as shown in Figs. 6(a) and (b). The square saturated magnetization curves without significant paramagnetic contribution indicate ideal magnetic nanoparticles rather than a core-shell magnetic structure[56]. This is further suggested by the almost absence of exchange bias effect. We have estimated the exchange-biased field, $H_{EB} = (H_1+H_2)/2$, from magnetization curves measured after FC in 1 T, where $H_1$ and $H_2$ are the positive and negative coercive fields, respectively. The observed exchange-biased field $H_{EB}$ is of negligible values, as presented in table 6, in the whole series which indicate the almost uniform spin structure with small surface disorder in the synthesized $Mg_{1-x}Ni_xFe_2O_4$ nanocrystals.

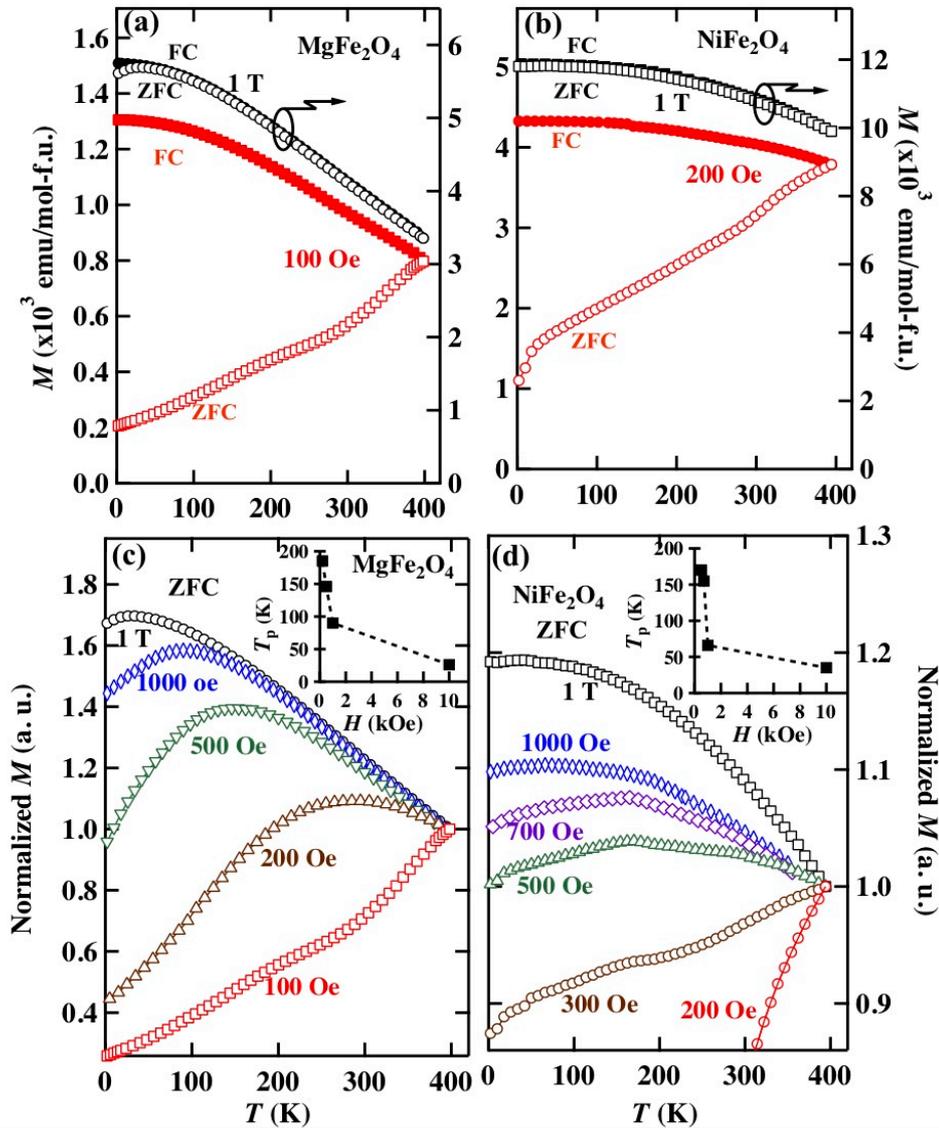

**Figure 5.** Temperature dependence of Fc and ZFC magnetization of $NiFe_2O_4$, (a) and (c), and $MgFe_2O_4$ nanoparticles, (b) and (d), measured at different magnetic fields. (c) and (d) show only the ZFC magnetization data and their insets show the field dependence of the peak temperature (blocking temperature) $T_p$.

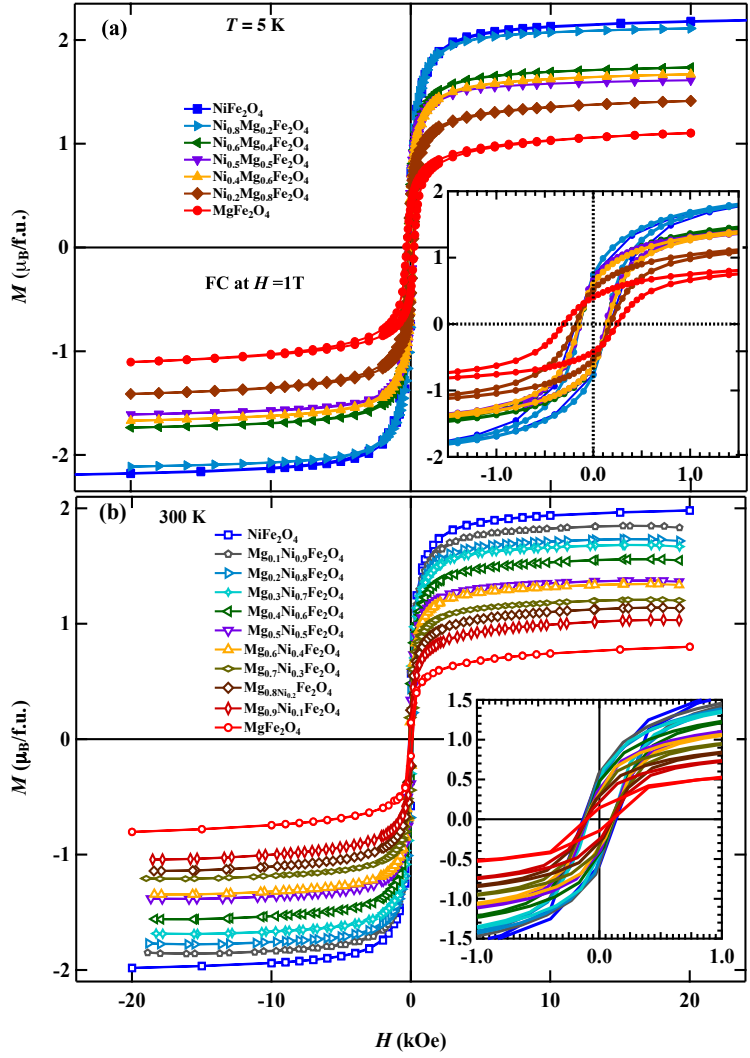

**Figure 6.** The magnetization curves (M vs. H) of $Mg_{1-x}Ni_xFe_2O_4$ nanoparticles measured at **(a)** 5 K and **(b)** 300 K after field cooling in a 1T magnetic field. The insets of (a) show magnification of the hysteresis loops around the zero field for all samples (right bottom) and the ZFC and FC curves for the $Mg_{0.5}Ni_{0.5}Fe_2O_4$ sample (top left). The inset of (b) shows magnification of the hysteresis loops around the zero field.

The spontaneous magnetization, $M_{sp}$, and remanence, $M_r$, as well as the coercive field, $H_c$, are estimated at 5 and 300 K and plotted against the Ni concentration ($x$) in Fig. 7. The values of magnetic moment per formula unit, $\mu_{st}$ ($\mu_B$), at room temperature are lower than those at 5 K, Fig. 7(a), which is normally due to the thermal agitation. While the enhanced magnetization with Ni-doping is attributed to the cations redistribution as well as replacing the nonmagnetic $Mg^{2+}$ with magnetic $Ni^{2+}$ ions those go to the *B* site and reinforce the majority spin. As the spontaneous magnetization is due to the difference between the magnetic moments in the two sublattices $^{IV}A$ and $^{VI}B$, we calculate the magnetic moment per formula unit using the cations distribution obtained for example from the Mössbauer spectroscopy, as, $\mu_{cal} = \mu^A - \mu^B$. Where $\mu^A$ and $\mu^B$ are the magnetic moment of constituent cations; $\mu(Mg^{2+}) = 0$, $\mu(Ni^{2+}) = 2$ and $\mu(Fe^{3+}) = 5$ $\mu_B$. The calculated values are presented in table 6 and shown as a function of Ni content in Fig. 7(a). It is clearly observed that the experimental values of $\mu_{sp}$ measured at 5 and 300 K, are comparable to theoretically calculated ones with same general linear behavior and $\mu_{cal}$ locates in between.

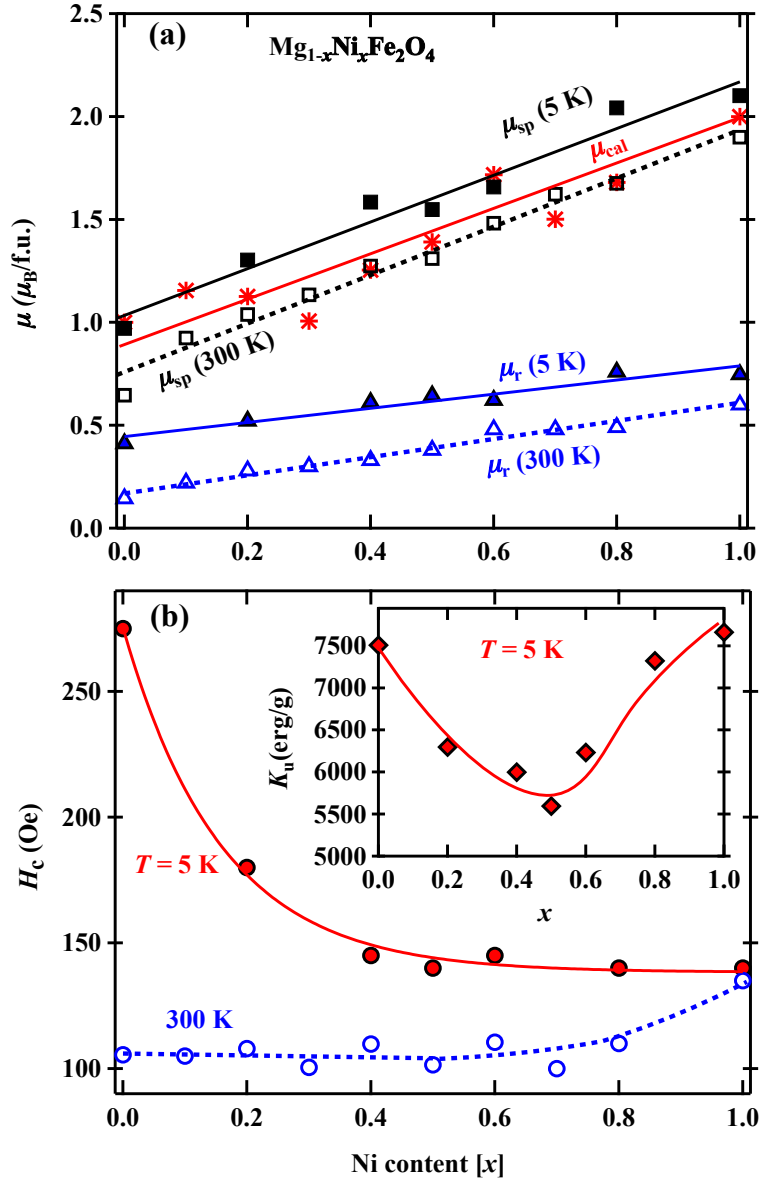

**Figure 7.** The Ni-content dependences of magnetic parameters, (a) spontaneous and remnant magnetic moments, $\mu_{sp}$ and $\mu_r$, measured at 5 K(closed) and 300 K (open) and the magnetic moment calculated based on the cations distribution results by Mossbauer spectroscopy at 78 K (stars). (b) the coercive field $H_c$ of $Mg_{1-x}Ni_xFe_2O_4$ nanoparticles measured at 5 and 300 K. The inset of (b) shows the evolution of anisotropy constant $K_u$ with Ni content.

Interestingly, the coercive field, $H_c$, measured at the low temperature of 5 K, which is much reliable, decreases with Ni doping, first rapidly up to $x = 0.5$ then maintains almost unchanged for higher Ni content. The anisotropy constant $K_u$, which is a substantial factor in the applications of soft magnetic materials, can be estimated according to the Stoner–Wohlfarth model for randomly oriented identical particles as[58,59],

$$K_u = \frac{H_c M_s}{0.96}, \qquad (4)$$

where $M_s$ is the saturation magnetization. The evolution of $K_u$ by Ni doping is estimated and presented for 5 K in the inset of Fig. 7(b) and table 6. It has a value of about $7.6 \times 10^3$ erg/g for the terminal compounds, $MgFe_2O_4$ and $NiFe_2O_4$, and shows a minimum value ($5.6 \times 10^3$ erg/g) for $Mg_{0.5}Ni_{0.5}Fe_2O_4$. In contrast to reported finite size effects[60], the nonmonotonic behavior of coercivity $H_c$ with its rapid decrease by Ni-doping up to $x = 0.5$ does not coincide the change in particle nor crystallite sizes, which excludes any role of reduced particle size enhance coercivity in our spinel ferrites of almost identical particle size. However, the $H_c$ behavior matches the trend of inversion factor shown in Fig. 2(b) with more Fe ions locates in the B sites for compositions below $x = 0.5$. The higher coercivity, and hence anisotropy field, for $MgFe_2O_4$ may be related to the higher majority-spin $Fe^{3+}$ ions number in the site with higher coordination number. With Ni-doping a rabid decrease in $Fe^{3+}$ ions in the B site may results in lowered $H_c$ until the almost complete inversion is reached with further doping. The progress in $K_u$ with minima at $x = 0.5$ is due to balanced contributions of the increased magnetization and decreased coercivity with Ni-doping. The reduced anisotropy factor with an enhanced magnetic moment implies further save of energy while magnetizing the material and suggests it as a candidate for applications of soft magnetic materials.

To confirm our view of the single-domain magnetism in our nanoparticles, we have estimated the single domain critical radius as[61],

$$R_{sd} = \frac{9\sqrt{A\,K_u}}{\mu_0 M_s^2}. \qquad (5)$$

Where $A$ is exchange stiffness that is given by $A = \frac{3\,K_B T_C}{z\,a}$, $K_B$ is the Boltzmann constant, $T_C$ is the Curie temperature, z is the effective number of the nearest neighbors and a is the lattice constant, all in the SI unit. Using $z \simeq 5$ and reported values of $T_C$[62], values of $R_{sd}$ are roughly estimated and presented in table 6. The single domain upper limit of size ranges in 30 – 40 nm for the synthesized ferrites with a smaller particle size (~ 20 – 30 nm) supports the single-domain magnetic behavior indicated by the temperature and field dependences described above.

The magnetic properties in $Mg_{1-x}Ni_xFe_2O_4$ solid-solution nanoparticles is not only influenced by the cations distribution, but the enhancement is also due to subsequent structural modifications after cations redistribution by Ni doping. The decrease in Me-Me and Me-O bond lengths and Me-O-Me angles breadth, as described above, altogether contribute to the observed enhancement of the magnetic properties. In view on the reported high electrical resistivity with lower Ni content[33–35], we introduce $Mg_{1-x}Ni_xFe_2O_4$ soft ferrite nanoparticles with tuneable magnetization and coercivity for high-frequency power applications. Further, our results shed lights on the correlation between the modified structural parameters by cations distribution and magnetic properties, introduced here for $Mg_{1-x}Ni_xFe_2O_4$ nanoparticles, which help in a demanded strategy to tune the magnetic properties of soft ferrite nanoparticles.

**Table 6:** Magnetic parameters of $Mg_{1-x}Ni_xFe_2O_4$ nanoparticles.

| $x$ ($Ni^{2+}$) | $T$ (K) | $M_{st}$ (emu/g) | $\mu_r$ ($\mu_B$) | $\mu_{st}$ ($\mu_B$) | $\mu_{sp}$ ($\mu_B$) | $\mu_{cal}$ ($\mu_B$)* | $H_c$ (Oe) | $H_{EB}$ (Oe) | $K_u$ (erg/g) | $T_C$ (K)† | $A$ ($\times 10^{-12}$ J/m) | $R_{sd}$ (nm)‡ |
|---|---|---|---|---|---|---|---|---|---|---|---|---|
| 0 | 5 | 26.21 | 0.41 | 1.102 | 0.9713 | 0.99 | 275 | 12.5 | 7509 | 625 | 5.16 | 42 |
|   | 300 | 19.08 | 0.15 | 0.802 | 0.64633 |  | 106 |  | 2097 |  |  |  |
| 0.1 | 300 | 24.54 | 0.22 | 1.032 | 0.92422 | 1.15 | 105 |  | 2684 | 638 | 5.27 | 43 |
| 0.2 | 5 | 33.59 | 0.52 | 1.412 | 1.3031 | 1.125 | 180 | 0 | 6297 |  |  |  |
|   | 300 | 27.02 | 0.28 | 1.136 | 1.0386 |  | 108 |  | 3040 |  |  |  |
| 0.3 | 300 | 28.59 | 0.30 | 1.202 | 1.1336 | 1.01 | 101 |  | 2993 | 725 | 7.19 | 41 |
| 0.4 | 5 | 39.71 | 0.61 | 1.669 | 1.5845 | 1.25 | 145 | 0 | 5998 |  |  |  |
|   | 300 | 31.81 | 0.33 | 1.337 | 1.2735 |  | 110 |  | 3637 |  |  |  |
| 0.5 | 5 | 38.38 | 0.65 | 1.613 | 1.548 | 1.39 | 140 | 5 | 5597 | 783 | 7.78 | 40 |
|   | 300 | 32.47 | 0.38 | 1.365 | 1.3099 |  | 102 |  | 3433 |  |  |  |
|   | 300 | 36.86 | 0.48 | 1.550 | 1.4823 | 1.72 | 110 |  | 4243 |  |  |  |
| 0.6 | 5 | 41.26 | 0.62 | 1.734 | 1.6579 | 1.72 | 145 | 2.5 | 6231 |  |  |  |
|   | 300 | 36.86 | 0.48 | 1.550 | 1.4823 |  | 110 |  | 4243 |  |  |  |
| 0.7 | 300 | 39.70 | 0.48 | 1.669 | 1.623 | 1.51 | 100 |  | 4136 | 813 | 8.08 | 31 |
| 0.8 | 5 | 50.21 | 0.76 | 2.111 | 2.0419 | 1.68 | 140 | 0 | 7322 |  |  |  |
|   | 300 | 40.87 | 0.49 | 1.718 | 1.6754 |  | 110 |  | 4684 |  |  |  |
| 0.9 | 300 | 43.99 | 0.58 | 1.849 | 1.7826 | 1.8 | 124 |  | 5659 |  |  |  |
| 1 | 5 | 52.54 | 0.75 | 2.209 | 2.1019 | 2 | 140 | 0 | 7663 | 870 | 10.83 | 29 |
|   | 300 | 47.18 | 0.60 | 1.983 | 1.8998 |  | 135 |  | 6635 |  |  |  |

* calculated using the cations distributions and the spontaneous magnetic moment ($\mu_{sp}$) estimated from Mossbauer data measured at 78 K.
† values reported for $Mg_{1-x}Ni_xFe_2O_4$ of comparable particle size are cited from ref. [62]
‡ roughly estimated according to equation (5) using average values of $K_u$ and $M_S$ measured at 5 K in SI units with the XRD-estimated density from ref. [37].

## 4. Conclusion

In this study, we have investigated the effect of $Ni^{2+}$ ions doping on the structure, cations distribution and magnetic properties of $Mg_{1-x}Ni_xFe_2O_4$ nanocrystals. In addition to the structural effects of $Ni^{2+}$ ions revealed by XRD and TEM such as lattice shrinkage, rabidly increased inversion in the spinel structure with low Ni-content ($x \leq 0.5$) and its sub-structural effects on average radii of ions and bond lengths between the tetrahedral $A$ and octahedral $B$ sites have been estimated by Rietveld refinements of the XRD patterns and Mössbauer spectroscopy. The magnetic moment evolution is corelated to the structural changes of cations-cations and cations-anions bond lengths and angles by Ni doping in $Mg_{1-x}Ni_xFe_2O_4$ nanoparticles. We found the coercive field measured at 5 K decreases dramatically from 275 to 140 Oe with increasing the Ni content up to $x = 0.5$ and remains almost unchanged above. The influence of $Ni^{2+}$ ions on the anisotropy factor that becomes lowest in $Mg_{0.5}Ni_{0.5}Fe_2O_4$ suggests these soft ferrites as candidates for high-frequency power applications.

**Acknowledgement**

Authors would like to thank Ms. Y. Uno for her help in the ICP-MS measurements and analysis. Authors also acknowledge the help of Mr. N. Sasaki and Mr. K. Kazumi with the TEM investigations.